\documentstyle[12pt]{article}
\input epsf
\newcommand{\be}{\begin{eqnarray}}
\newcommand{\ee}{\end{eqnarray}}
\newcommand{\la}[1]{\label{#1}}
\newcommand{\eq}[1]{eq.~(\ref{#1})}

 \def\Tr{\mbox{Tr}}
 
\def\Dirac1#1{#1\hskip-2pt/}
\def\Dirac#1{#1\hskip-6pt/}
\def\ns{\Dirac n}
\def\dd{\Dirac \partial}
  \def\beq{\begin{equation}}
  \def\eeq{\end{equation}}
\begin{document}
%
%
\rightline{RUB-TPII-8/97}
\vspace{.3cm}
\begin{center}
\begin{large}
{\bf Off-forward quark distributions of the nucleon \\
in the large $N_c$ limit}
\\
\end{large}
\vspace{1.4cm}
{\bf V. Yu.\ Petrov}, {\bf P.V.\ Pobylitsa},
{\bf M.V.\ Polyakov}\footnote{e-mail:
maximp@hadron.tp2.ruhr-uni-bochum.de}\\[0.2 cm]
{\em Petersburg Nuclear Physics Institute, Gatchina,
St.Petersburg 188350, Russia}\\[0.3 cm]
{\bf I.\ B\"ornig}, {\bf K. Goeke}, {\bf C.\ Weiss}
\\[0.2 cm]
{\em Institut f\"ur Theoretische Physik II,
Ruhr--Universit\"at Bochum, \\ D--44780 Bochum, Germany} \\
\end{center}
\vspace{1.5cm}
\begin{abstract}
\noindent
We study the off-forward quark distributions (OFQD's)
in the chiral quark-soliton model of the nucleon. This model
is based on the large-$N_c$ picture of the nucleon as a soliton of the
effective chiral lagrangian and allows to
calculate the leading twist quark-- and antiquark distributions at a
low normalization point. We demonstrate the consistency of the approach
by checking various sum rules for the OFQD's. We present numerical estimates
of the isosinglet distribution $H(x,\xi,\Delta^2)$. In contrast to other
approaches we find a strong qualitative dependence on the longitudinal
momentum transfer, $\xi$. In particular, $H(x,\xi,\Delta^2)$ as a
function of $x$ exhibits fast crossovers at $|x| = \xi/2$.
Such behaviour could lead to a considerable
enhancement of the deeply--virtual Compton scattering cross section.
\end{abstract}
%
%
%
%
\newpage

\renewcommand{\theequation}{\arabic{section}.\arabic{equation}}

\setcounter{equation}{0}

\section{Introduction}

Although a large portion of quantitative information about strong
interactions is contained in the well-known parton distribution functions
of the nucleon, one should not forget that they
provide us with still far from complete knowledge of the structure of the
nucleon. Recently, a new type of parton distributions has attracted
considerable interest, the so called off--forward parton distributions
(OFPD's), which are generalizations simultaneously of the usual parton
distributions and of the elastic nucleon form factors. Taking the $n$--th
moment of the OFPD's one obtains the form factors ({\em i.e.}, non-forward
matrix elements) of the spin--$n$, twist--two quark and gluon operators.
On the other hand, in the forward limit the OFPD's reduce to the usual
quark, antiquark and gluon distributions. In other words, the OFPD's
interpolate between the traditional inclusive (parton distributions) and
exclusive (form factors) characteristics of the nucleon and thus
provide us with a considerable new amount of information on nucleon
structure.

The OFPD's are not accessible in standard inclusive measurements. They can,
however, be measured in deeply--virtual Compton scattering (DVCS)
\cite{Ji1,Rad,Ji2,Ji3,Rad3} and in hard exclusive electroproduction of mesons
\cite{Rad2,CFS,Rad3}. A firm theoretical basis for the QCD analysis of
DVCS and hard exclusive electroproduction of mesons in terms of off--forward
parton distributions in the nucleon is provided by the recently proven
factorization theorems for these processes \cite{Rad2,CFS,Rad3}.
A quantitative description of these classes of processes requires not only
knowledge of the perturbative evolution of the OFPD's
\cite{Ji2,Rad2,Rad3,evh}, but also non-perturbative information
in the form of the OFPD's at some initial normalization point. Thus, a
computation of the OFPD's at a low normalization point in a realistic
model of the nucleon is of great importance. While the usual
parton distributions have been measured in
a variety of different experiments and can be confronted
with model calculations {\em a posteriori}, in the case of the off--forward
distributions the situation is rather opposite: Here, model calculations
of the distributions at a low normalization point are required to
determine the very feasibility of measuring the OFPD's, say, in DVCS at
experimental energies.

Recently it was shown that the chiral quark--soliton model of the nucleon
\cite{DPP}, which is based on the instanton model of the QCD
vacuum, provides a framework for a succesful calculation of the nucleon
parton distributions, both unpolarized and polarized, at a low normalization
point \cite{DPW,DPPPW,PP,DPPPW2}. This is a quantum field--theoretical
description of the nucleon, with explicit quark degrees of freedom, which
allows an unambiguous identification of the quark as well as antiquark
distributions in the nucleon. We were able to demonstrate that all
general properties of the quark and antiquark distributions (positivity,
sum rules, Soffer inequalities, {\em etc.})
are correctly realized in this description. At the same time,
the chiral quark--soliton model gives a good account of a variety of
static properties of the nucleon and other baryons, and of baryon form
factors (for a review, see ref.~\cite{Review}). Given the success
in describing both nucleon form factors and the usual parton
distribution functions, this approach seems well posed for a theoretical
investigation of the off--forward parton distributions of the nucleon
at a low normalization point.

Recently, Ji {\it et al.} have studied the off--forward quark distributions
in the bag model \cite{Ji3}, providing the first model estimates of these
quantities. However, bag models
encounter severe problems in applications to parton distribution functions.
It is usually assumed that the three quarks in the bag give
rise only to quark distributions; however, they inevitably produce also
an antiquark distribution with {\em negative} sign. This circumstance
is not often emphasized. To overcome this problem one would need to add the
contribution to the structure function arising from the forces which bind
the quarks --- in the case of the bag model, from the bag surface.
Unfortunately, the latter is not described in terms of fields.
In contrast to the bag
model our approach ensures the positivity of both quark and antiquark
distributions, because the forces which bind the quarks in the
nucleon are described consistently in terms of fields.
Below we shall see that in the case of off-forward quark distributions
(OFQD's) the field-theoretic description leads to characteristic crossovers
of the OFQD's at $x = \pm\xi/2$, as opposed to the bag model result,
which is smooth at these points.

In this paper we study the off-forward quark and antiquark
distributions in the chiral quark--soliton model of the nucleon.
In section \ref{section_QCD_def} we outline the QCD definition
of the OFPD's and their basic properties. Section \ref{section_model}
gives a brief introduction to the chiral quark soliton model of the
nucleon. In section \ref{section_OFQD_in_model} we derive the
expressions for the OFQD's in the chiral quark soliton model
and show that the sum rules are satisfied within the model. We
discuss the contributions to the OFQD's from the bound--state level
of quarks as well as from the Dirac continuum, whose presence
is a consequence of the field--theoretic character of this description of
the nucleon. We show that the latter is crucial near $x = \pm\xi/2$
and lead to sharp crossovers of the OFQD's. Numerical estimates
for the singlet unpolarized distribution $H(x,\xi ,\Delta ^2)$ are
presented in section \ref{section_numeric}. A summary and conclusions
are given in section \ref{section_summary}.

\section{QCD definition of off-forward quark distributions}
\label{section_QCD_def}
In QCD the off-forward parton distributions are defined through
nondiagonal matrix elements of product of quark fields at light--cone
separation. Here and in the following, we shall use the notations
of refs.~\cite{Ji1,Ji2}:
\be
\nonumber
\int \frac{d\lambda }{2\pi }e^{i\lambda x}\langle P^{\prime }|\bar \psi
(-\lambda n/2){\Dirac n}
 \psi (\lambda n/2)|P\rangle &=& H(x,\xi ,\Delta^2) \;
\bar U(P^{\prime }) \; \ns \; U(P) \\
&+& \frac 1{2M_N} \; E(x,\xi ,\Delta^2) \;
\bar U(P^{\prime }) \; i\sigma ^{\mu \nu }\, n_\mu \; \Delta _\nu U(P) .
\nonumber \\
\label{E-H-QCD-2}
\ee
Here $n_\mu$ is a light-cone vector,
\be
n^2 &=& 0, \hspace{2cm} n\cdot (P + P') \;\; = \;\; 2,
\label{n-normalization}
\ee
$\Delta$ the four--momentum transfer,
\begin{equation}
\Delta = P^{\prime }-P , \label{Delta-def}
\end{equation}
$M_N$ denotes the nucleon mass, and $U(P)$ is a standard Dirac
spinor. The off-forward quark distributions, $H(x,\xi ,\Delta ^2)$ and
$E(x,\xi ,\Delta ^2)$, are regarded as functions of the variable
$x$, the square of the four--momentum transfer, $\Delta^2$,
and its longitudinal component
\begin{equation}
\xi =-(n\cdot \Delta ) . \label{xi-def}
\end{equation}
\par
In the forward case, $P = P'$, both $\Delta$ and $\xi$ are zero, and
the second term on the r.h.s.\ of eq.(\ref{E-H-QCD-2}) disappears. The
function $H$ becomes the usual parton distribution function,
\be
H(x, \xi = 0, \Delta ^2 = 0) &=& q(x) .
\label{forward_limit}
\ee
On the other hand, taking the first moment of eq.(\ref{E-H-QCD-2})
one reduces the operator on the l.h.s. to the local vector current.
The dependence of $H$ and $E$ on $\xi$ disappears, and the functions
reduce to the usual electric and magnetic form factors of the nucleon,
\be
\int_{-1}^1 dx \; H(x, \xi , \Delta ^2) &=& F_1 (\Delta^2 ), \\
\int_{-1}^1 dx \; E(x, \xi , \Delta ^2) &=& F_2 (\Delta^2 ).
\ee
Taking higher moments of the distribution functions one obtains the form
factors of the twist--2, spin--$n$ operators.

\section{Chiral quark--soliton model of the nucleon}
\label{section_model}

Recently a new approach to the calculation of quark distribution functions
has been developed \cite{DPPPW} within the context of the chiral
quark-soliton model of the nucleon \cite{DPP}. In this paper we apply this
approach to the calculation of OFQD's. It is essentially based on the
$1/N_c$ expansion.  Although in reality the number of colours $N_c=3$,
the academic limit of large $N_c$ is known to be a useful guideline. At
large $N_c$ the nucleon is heavy and can be viewed as a classical
soliton of the pion field \cite{Witten,ANW}.  In this paper we work
with the effective chiral action given by the functional integral over
quarks in the background pion field \cite{DE,DSW,DP}:

\begin{eqnarray}
\exp \left( iS_{{\rm eff}}[\pi (x)]\right) =\int D\psi D\bar \psi \;\exp
\left( i\int d^4x\,\bar \psi (i\dd -MU^{\gamma _5})\psi \right) ,
\nonumber
\end{eqnarray}
\begin{equation}
U\;=\;\exp \left[ i\pi ^a(x)\tau ^a\right] ,\hspace{1cm}U^{\gamma
_5}\;=\;\exp \left[ i\pi ^a(x)\tau ^a\gamma _5\right] \;=\;\frac{1+\gamma _5}
2U+\frac{1-\gamma _5}2U^{\dagger }.  \label{FI}
\end{equation}
Here $\psi $ is the quark field, $M$ the effective quark mass, which is
due to the spontaneous breakdown of chiral symmetry (generally
speaking, it is momentum dependent), and $U$ is the $SU(2)$ chiral pion
field. The effective chiral action given by (\ref{FI}) is known
to contain automatically the Wess--Zumino term and the four-derivative
Gasser--Leutwyler terms, with correct coefficients.
Equation (\ref{FI}) has been
derived from the instanton model of the QCD vacuum \cite{DP,DP1}, which
provides a natural mechanism of chiral symmetry breaking and enables one to
express the dynamical mass $M$ and the ultraviolet cutoff intrinsic in (\ref
{FI}) through the $\Lambda _{QCD}$ parameter. The ultraviolet
regularization of the effective theory is provided by the specific momentum
dependence of the mass, $M(p^2)$, which drops to zero for momenta of
order of the inverse instanton size in the instanton vacuum,
$1/\rho \sim 600\, {\rm MeV}$. For simplicity we shall neglect this momentum
dependence in the general discussion; it will be taken into account again
in the theoretical analysis and in the numerical estimates later.

An immediate application of the effective chiral theory (\ref{FI}) is the
quark-soliton model of baryons of ref.~\cite{DPP}, which is in the spirit of
the earlier works~\cite{KaRiSo,BiBa}. According to this model nucleons can
be viewed as $N_c$ ``valence" quarks bound by a self-consistent pion
field (the ``soliton") whose energy coincides with the aggregate
energy of the quarks of the negative-energy Dirac continuum. Similarly to
the Skyrme model large $N_c$ is needed as a parameter to justify the use of
the mean-field approximation; however, the $1/N_c$--corrections can be
--- and, in some cases, have been --- computed \cite{Review}.

Let us remind the reader how the nucleon is described in the effective
low-energy theory (\ref{FI}). Integrating out the quarks in (\ref{FI}) one
finds the effective chiral action,
\begin{equation}
S_{{\rm eff}}[\pi ^a(x)]=-N_c\,\mbox{Sp}\log D(U)\,,\hspace{1cm}
D(U)\;\;=\;\;i\partial _0-H(U),  \label{SeffU}
\end{equation}
where $H(U)$ is the one-particle Dirac hamiltonian,
\begin{equation}
H(U)=-i\gamma ^0\gamma ^k\partial _k+M\gamma ^0U^{\gamma _5}\,,  \label{hU}
\end{equation}
and $\mbox{Sp}\ldots $ denotes the functional trace.
For a given time-independent pion field $U=\exp(i\pi^a({\bf x})\tau^a)$ one
can determine the spectrum of the Dirac hamiltonian,
\begin{equation}
H\Phi_n = E_n \Phi_n.  \label{Dirac-equation}
\end{equation}
It contains the upper and lower Dirac continua (distorted by the presence of
the external pion field), and, in principle, also discrete bound-state
level(s), if the pion field is strong enough. If the pion field has unity
winding number, there is exactly one bound-state level which travels all the
way from the upper to the lower Dirac continuum as one increases the spatial
size of the pion field from zero to infinity \cite{DPP}. We denote the
energy of the discrete level as $E_{{\rm lev}},\;\;-M\leq E_{{\rm lev}}\leq
M $. One has to occupy this level to get a non-zero baryon number state.
Since the pion field is colour blind, one can put $N_c$ quarks on that level
in the antisymmetric state in colour.

The limit of large $N_c$ allows us to use the mean-field approximation to
find the nucleon mass. To get the nucleon mass one has to add
$N_cE_{{\rm lev}}$ and the energy of the pion field.
Since the effective chiral lagrangian
is given by the determinant (\ref{SeffU}) the energy of the pion field
coincides exactly with the aggregate energy of the lower Dirac continuum,
the free continuum subtracted. The self-consistent pion field is thus found
from the minimization of the functional \cite{DPP}

\begin{equation}
M_N = \min_U \; N_c\left\{E_{{\rm lev}}[U] \; + \;
\sum_{E_n<0}(E_n[U]-E_n^{(0)})\right\}.  \label{nm}
\end{equation}
{From} symmetry considerations one looks for the minimum in a hedgehog ansatz:
\begin{equation}
U_c({\bf x}) \; = \; \exp\left[i\pi^a({\bf x})\tau^a\right] \; = \;
\exp\left[i n^a \tau^a P(r)\right], \hspace{1cm} r \; = \; |{\bf x}|,
\hspace{1cm} {\bf n} \; = \; \frac{{\bf x}}{r} ,  \label{hedge}
\end{equation}
where $P(r)$ is called the profile of the soliton.

The minimum of the energy (\ref{nm}) is degenerate with respect to
translations of the soliton in space and to rotations of the soliton field
in ordinary and isospin space. For the hedgehog field (\ref{hedge}) the two
rotations are equivalent.
The projection on a nucleon state with
given spin ($S_3$) and isospin ($T_3$) components
is obtained by integrating over all spin-isospin rotations,
$R$ \cite{ANW,DPP},
\be
\langle S=T,S_3,T_3|\ldots| S=T,S_3,T_3\rangle &=& \int
dR\;\phi^{\ast\;S=T}_{S_3T_3}(R) \; \ldots \; \phi^{S=T}_{S_3T_3}(R)\,.
\label{spisosp}
\ee
Here $\phi^{S=T}_{S_3T_3}(R)$ is the rotational wave function of the nucleon
given by the Wigner finite-rotation matrix \cite{ANW,DPP}:
\be
\phi _{S_3T_3}^{S=T}(R) &=& \sqrt{2S+1}(-1)^{T+T_3}D_{-T_3,S_3}^{S=T}(R).
\label{Wigner}
\ee
Analogously, the projection on a nucleon state with given momentum
${\bf P}$ is obtained by integrating over all shifts, ${\bf X}$, of the
soliton,
\be
\langle {\bf P^\prime}|\ldots|\ {\bf P}\rangle
&=& \int d^3{\bf X}\;e^{i({\bf P^\prime-P})\cdot{\bf X}}\; \ldots
\label{totmom}
\ee
\section{Off-forward quark distributions in the chi\-ral
quark--so\-li\-ton
model}
\label{section_OFQD_in_model}
We now turn to the calculation of the off-forward quark distributions in
the chiral quark--soliton model. This description of the nucleon
is based on the $1/N_c$--expansion. At large $N_c$ the nucleon is
heavy --- its mass is $O(N_c)$.
Thus for the large--$N_c$ nucleon
eq.~(\ref{E-H-QCD-2}) simplifies as follows:
\be
\int \frac{d\lambda }{2\pi }e^{i\lambda x}\langle P^{\prime },S_3^{\prime
}|\bar \psi (-\lambda n/2){\Dirac n}
\psi (\lambda n/2)|P,S_3\rangle
&=&2\delta
_{S_3^{\prime }S_3}H(x,\xi ,\Delta ^2)
\label{E-H-QCD-3} \\
 &-&\frac i{M_N}\epsilon ^{3jk}\Delta ^j
(\sigma ^k)_{S_3^{\prime}S_3}
E(x,\xi ,\Delta ^2)  \, ,
\nonumber
\ee
where $S_3, S_3'$ denote the projections of the nucleon spin.
\par
Before computing the quark distribution functions we must determine the
parametric order  in $1/N_c$ of the kinematical variables involved.
Generally, when describing parton distributions in the large--$N_c$ limit,
$x \sim 1/N_c$, since the nucleon momentum is distributed
among $N_c$ quarks. Furthermore, as in the calculation of nucleon form
factors we
consider momentum transfers $\Delta^2 \sim N_c^0$; hence, in particular,
$\xi \sim 1/N_c$, so that $\xi$ is of the same parametric order as
$x$.

The calculation of the off--forward parton distributions proceeds in much
the same way as that of the usual parton distributions \cite{DPPPW}.
One expands the quark fields in eq.(\ref{E-H-QCD-3}) in the basis of
quark single--particle wave functions in the background pion field,
eq.(\ref{Dirac-equation}). The nucleon matrix element is then obtained
by summing over all occupied single--particle states (the bound--state
level and the negative--energy continuum) and projecting on nucleon
states with definite spin and momenta by integrating over collective
coordinates of the soliton field with appropriate
wave functions, {\em cf.}\ eqs.(\ref{spisosp}, \ref{totmom}).
When integrating over soliton rotations one must consider separately
flavour singlet and nonsinglet matrix elements. In the
flavour singlet case the integral over soliton rotations is trivial,
and one obtains
\be
\nonumber
\lefteqn{
\int
\frac{d\lambda }{2\pi } e^{i\lambda x} \;
\sum\limits_f \; \langle P^{\prime
},T^{\prime }=S^{\prime },T_3^{\prime },S_3^{\prime }| \;
\bar \psi _f(-\lambda
n/2)\ns\psi _f(\lambda n/2) \; |P,T=S,T_3,S_3\rangle } && \\
\nonumber
&=&
\frac{N_c M_N^2}\pi \;
\delta _{TT^{\prime }}\delta _{SS^{\prime
}}\delta _{T_3T_3^{\prime }}\delta _{S_3S_3^{\prime }} \;
\int dz^0\int d^3 {\bf X} \;
\exp [i\mbox{{\boldmath $\Delta$}}\cdot {\bf X}] \\
&&\times \sum\limits_{{\scriptstyle}\atop {\scriptstyle {\rm occup.}}}\exp
\{iz^0[(x+\xi/2)M_N-E_n]\} \;
\Phi _n^{\dagger }({\bf X})\gamma^0\ns
\Phi _n({\bf X}-z^0{\bf e}_3) \,,
\label{for-H}
\ee
where ${\bf e}_3$ is the unit vector in the third direction. This equation
was derived in \cite{DPPPW} for the case $P^{\prime }=P$; the generalization
to $P^{\prime }\ne P$ is straightforward. [Here and in the analogous
expressions below it is understood that one subtracts the
sums over levels of the vacuum hamiltonian $(U = 1)$.] For the flavour
nonsinglet part, on the other hand, one has
\be
\nonumber
\lefteqn{\int \frac{d\lambda }{2\pi }e^{i\lambda x} \;
\langle P^{\prime
},T^{\prime }=S^{\prime },T_3^{\prime },S_3^{\prime }| \; \bar \psi
(-\lambda n/2)\tau ^a\ns\psi (\lambda n/2) \; |P,T=S,T_3,S_3\rangle } &&
\\
\nonumber
&=&
\frac{N_cM_N^2}\pi \int dz^0\int d^3 {\bf X} \;  \exp [i
\mbox{{\boldmath $\Delta$}}\cdot {\bf X}] \;
\int dR\,\phi _{T_3^{\prime }S_3^{\prime }}^{T^{\prime }=S^{\prime
}*}(R)
\phi _{T_3S_3}^{T=S}(R)\;\frac{1}{2} \Tr(R^{\dagger }\tau ^aR\tau ^b)\\
&&\times
\sum\limits_{{\scriptstyle}\atop {\scriptstyle {\rm occup.}}}\exp
\{iz^0[(x+\xi/2)M_N-E_n]\} \;
\Phi _n^{\dagger }({\bf X})\tau
^b\gamma ^0\ns\Phi _n({\bf X}-z^0{\bf e}_3)  \, .
\label{for-E}
\ee
Here $\phi _{T_3S_3}^{T=S}(R)$ are the rotational wave functions of the
soliton, eq.(\ref{Wigner}). The integral over the soliton orientation
matrix $R$ can be computed,
\begin{equation}
\int dR\,\phi _{T_3^{\prime }S_3^{\prime }}^{\frac 12*}(R)\phi
_{T_3S_3}^{\frac 12}(R)\;\frac 12\Tr(R^{\dagger }\tau ^aR\tau ^b)
\; = \; -\frac
13(\tau ^a)_{T_3^{\prime }T_3}(\sigma^b)_{S_3^{\prime }S_3} \, .
\label{rot_isovector}
\end{equation}
Comparing eq.(\ref{for-H}) and eqs.(\ref{for-E}, \ref{rot_isovector})
with eq.(\ref{E-H-QCD-3}) we immediately see that in the leading order of
the $1/N_c$ expansion only the flavour singlet part of
$H(x,\xi ,\Delta ^2)$ and the flavour--nonsinglet
part of $E(x,\xi ,\Delta ^2)$ are non-zero. They are given, respectively,
by
\be
\lefteqn{
\sum\limits_fH_f(x,\xi ,\Delta ^2)  \;\; = \;\;
\frac{N_cM_N}{2\pi }\int dz^0\int
d^3{\bf X} \; \exp [i\mbox{{\boldmath $\Delta$}}\cdot {\bf X}] }
\nonumber \\
&&\times \sum\limits_{{\scriptstyle}\atop {\scriptstyle {\rm occup.}}}\exp
\{iz^0[(x+\xi/2)M_N-E_n]\} \; \Phi _n^{\dagger }({\bf X}
)(1+\gamma ^0\gamma ^3)\Phi _n({\bf X}-z^0{\bf e}_3)\, ,
\label{H-singlet-general} \\
\lefteqn{\epsilon ^{3jk}\Delta ^jE^{(3)}(x,\xi ,\Delta ^2)
\;\; = \;\; -\frac{iN_cM_N^2}{3\pi} \int dz^0\int d^3{\bf X} \; \exp [i
\mbox{{\boldmath $\Delta$}}\cdot {\bf X}] } &&
\nonumber \\
&&\times \sum\limits_{{\scriptstyle}\atop {\scriptstyle {\rm occup.}}}\exp
\{iz^0[(x+\xi/2)M_N-E_n]\} \; \Phi _n^{\dagger }({\bf X})\tau
^k(1+\gamma ^0\gamma ^3)\Phi _n({\bf X}-z^0{\bf e}_3)\, .
\label{E-nonsinglet-general}
\ee
The isovector part of $H(x,\xi ,\Delta ^2)$ and the isosinglet part
of $E(x,\xi ,\Delta ^2)$ appear only in the next--to--leading order
of the $1/N_c$--expansion, {\em i.e.}, after taking into account
the finite angular velocity of the soliton rotation.
\par
Before going ahead with the evaluation of the
expressions eqs.(\ref{H-singlet-general}, \ref{E-nonsinglet-general})
we would like to demonstrate that the two limiting cases
of the off--forward distributions
--- usual parton distributions and elastic form factors --- are
correctly reproduced within the chiral quark--soliton model.
Taking in eq.(\ref{H-singlet-general}) the forward limit,
$\Delta\rightarrow 0$, one recovers the formula for the usual
singlet (anti--) quark distributions in our model which was
obtained in ref.\cite{DPPPW}. Thus the forward limit,
eq.(\ref{forward_limit}), is reproduced. On the other hand, integrating
eqs.(\ref{H-singlet-general},\ref{E-nonsinglet-general}) over $-1\le
x\le 1$ one obtains (up to corrections parametrically small in $1/N_c$)
the expressions for the electromagnetic formfactors of the nucleon
derived in ref.~\cite{DPP}:
\be
\nonumber
\int_{-1}^1 dx\ \sum\limits_fH_f(x,\xi ,\Delta ^2) &=&
N_c \int d^3{\bf X}\exp[i\mbox{{\boldmath $\Delta$}}\cdot {\bf X}]
\sum\limits_{{\scriptstyle}\atop {\scriptstyle {\rm occup.}}}
\Phi _n^{\dagger }({\bf X})\Phi _n({\bf X})
\nonumber \\
&=& F_1^{\rm (T=0)}(\Delta^2)\, ,
\label{H-singlet-sum-rule}\\
\nonumber
\int_{-1}^1 dx\  E^{(3)}(x,\xi ,\Delta ^2) &=&
-\frac{i N_cM_N}{\Delta^2} \varepsilon_{3ik}\Delta^i
\int d^3{\bf X}\exp[i\mbox{{\boldmath $\Delta$}}\cdot {\bf X}]
\sum\limits_{{\scriptstyle}\atop {\scriptstyle {\rm occup.}}}
\Phi _n^{\dagger }({\bf X})\gamma^0\gamma^3\tau^k
\Phi _n({\bf X}) \\
&=& F_2^{(\rm T=1)}(\Delta^2)\, .
\label{E-nonsinglet-sum-rule}
\ee
The electromagnetic formfactors computed in the chiral quark
soliton model on the basis of these formulas compare very well
with the experimentally measured ones up to momenta of order
$\Delta^2\sim 1$~GeV$^2$ \cite{Review,CGPG}.

Eqs.~(\ref{H-singlet-general},\ref{E-nonsinglet-general}) express
the OFPD's as a sum over quark single--particle levels in the
soliton field. This sum runs over {\it all} occupied levels, including
both the discrete bound--state level and the negative Dirac continuum.
We remind the reader that in the case of usual parton distributions
it was demonstrated that in order to ensure the positivity of the antiquark
distributions it is essential to take into account the contributions
of {\it all} occupied levels of the Dirac Hamiltonian \cite{DPPPW}.
The so-called ``valence level aproximation'' for structure
functions in the chiral quark-soliton model advocated in
\cite{GamWeiRei} leads to unacceptable {\it negative}
antiquark distributions. We shall see below that also in the
off-forward case the contribution of the Dirac continuum drastically
changes the shape of the distribution function, leading to
characteristic crossovers of $H(x,\xi,\Delta)$ at $|x|=\xi/2$.
\par
We shall now compute the contributions
of the discrete bound--state level and the negative Dirac continuum
to eqs.~(\ref{H-singlet-general},\ref{E-nonsinglet-general}).
We focus here on the isosinglet distribution $H(x,\xi,\Delta)$; the
discussion and the expressions derived below can be easily generalized
to the case of other OFQD's.
\par
The contribution of the discrete bound--state level to
eq.~(\ref{H-singlet-general}) can be computed using the expressions
given in the Appendix. The result is shown in Fig.~1 for the forward case
and Fig.~2 for a non-zero momentum transfer. Being taken by itself this
contribution resembles the shape of $H(x,\xi,\Delta^2)$ obtained in
the bag model \cite{Ji3}.
\par
To calculate the contribution of the Dirac continuum to
eq.~(\ref{H-singlet-general}) we resort to an appro\-xima\-tion
which proved to be very successful in the computation of usual parton
distributions, the so--called interpolation formula \cite{DPPPW}.
One first expresses the continuum contribution as a functional trace
involving the quark propagator in the background pion field. The
quark propagator can then be expanded in powers of the formal parameter
$\partial U/(-\partial^2+M^2)$, which becomes small
in three limiting cases: {\em i}) low momenta, $|\partial U|\ll M$,
{\em ii}) high momenta, $|\partial U|\gg M$, {\em iii}) any momenta but
small pion fields, $|\log U| \ll 1$. One may therefore expect that this
approximation has good accuracy also in the general case. As was shown in
refs.~\cite{DPPPW,DPPPW2} for usual parton distributions this approximation
preserves the positivity of the antiquark distributions and all sum rules;
moreover, it gives results very close to those obtained by exact numerical
diagonalization of the Dirac hamiltonian and summation over the
negative--energy levels.

To derive the interpolation formula for the isosinglet distribution
$H(x,\xi ,\Delta ^2)$ (the expressions can easily be generalized to
the case of isovector $E^{(3)}(x,\xi ,\Delta ^2)$) we proceed
in analogy to the case of usual parton distributions \cite{DPPPW}
and rewrite the continuum contribution to eq.~(\ref{H-singlet-general})
as
\be
\lefteqn{\left[ \sum\limits_fH_f(x,\xi ,\Delta ^2)\right] _{\rm cont}
\;\; = \;\; \frac{N_c M_N}{2\pi T }
\; \mbox{Im} \; \int dz^0 \; e^{i z^0 x M_N} }
\nonumber \\
&&\times
\mbox{Sp}\left\{ (v_\mu \gamma ^\mu )\exp [i\mbox{{\boldmath $\Delta$}}
\cdot {\bf X}+z^0(v_\mu \partial ^\mu )]
\frac {1}{\left[ i\gamma ^\mu
\partial _\mu -MU^{\gamma _5}+i0\right]}\right\} \; - \;
(U\rightarrow 1) \, .
\label{Feynman-representation}
\ee
Here we have introduced a dimensionless light-like vector,
$v_\mu=(1,0,0,1)$, and $T$ is the time interval which is canceled
by a corresponding factor arising from the functional trace.
We have written here explicitly the vacuum subtraction term.
Expanding now in $\partial U/(-\partial^2+M^2)$ and evaluating the
the functional trace in momentum space one obtains in leading order:
\be
\nonumber
\lefteqn{\left[ \sum\limits_fH_f(x,\xi ,\Delta ^2)\right]_{\rm cont} }
&& \nonumber \\
&=& -2 M_N N_c \; \mbox{Im} \; \int \frac{d^3k}{(2\pi )^3}\int
\frac{d^4p}{(2\pi )^4} \;
\delta
\left[ \Bigl(x-\frac \xi 2 \Bigr) M_N-v\cdot p\right]
\nonumber \\
&&\times
\frac{M(p^2)}{(p^2-M_0^2+i0)}\, \frac{M((p-k)^2)}{(p-k)^2-M_0^2+i0}
(k\cdot v) \; \Tr_{\rm fl.}
\left\{ \tilde U({\bf k-}\mbox{{\boldmath $\Delta$}})
[\tilde U({\bf k})]^{+}\right\}
\nonumber \\
&+&\Biggl( \xi \to -\xi, \;\; \mbox{{\boldmath $\Delta$}}
\to -\mbox{{\boldmath $\Delta$}}
\Biggr)\, ,
\label{H-1-sym-res-mp}
\ee
where the Fourier transform of the soliton field is defined as

\begin{equation}
\nonumber
\tilde U({\bf k)\equiv }\int d^3{\bf x\,}e^{-i {\bf k} \cdot {\bf x} }\,
\bigl[U({\bf x})-1\bigr] \,.
\end{equation}
In eq.~(\ref{H-1-sym-res-mp}) we have re-instated the momentum
dependence of the constituent quark mass, $M(p^2)$, which cuts the
loop momentum $p$ and thus regularizes the UV divergence; by
$M_0$ we denote the value of the mass at zero momentum, $M(0)$.
We have neglected in eq.~(\ref{H-1-sym-res-mp}) the momentum dependence
of the quark masses appearing in the denominators; the masses
standing in the denominators do not play the role of an UV regulator
and their momentum dependence is not essential.
\par
In ref.\cite{DPPPW} the continuum contribution to the quark distribution
functions was computed by regularizing the loop integrals with
a relativistic Pauli--Villars cutoff. One may argue that this
regularization mimics the effect of the momentum dependence of the
constituent quark mass. [We shall soon see under which conditions
this assumption is valid.] Let us evaluate the continuum contribution to
the OFQD, eq.~(\ref{H-1-sym-res-mp}), also with a Pauli--Villars cutoff,
neglecting the momentum dependence of the masses in the numerator.
One obtains
\be
\lefteqn{ \left[ \sum\limits_fH_f(x,\xi ,\Delta ^2)\right] _{\rm cont}
\;\; = \;\;
\frac{N_cM_NM_0^2}{2\pi }\; \mbox{sign}\left( x-\frac \xi 2\right) }
\nonumber
\\
\nonumber
&&\times
\int \frac{d^3k}{(2\pi )^3} \; \theta \left\{ -\left( x-\frac \xi
2\right) M_N\left[ \left( x-\frac \xi 2\right) M_N-k^3\right] \right\}
\mbox{Re\thinspace Tr}_{fl}\left\{ \tilde U({\bf k-}
\mbox{{\boldmath $\Delta$}} )
\left[ \tilde U(
{\bf k})\right] ^{\dagger }\right\}\\
\nonumber
&&\times \int \frac{d^2p_{\perp }}{(2\pi )^2}
\frac{1}{|p_{\perp}|^2 + M_0^2
- \frac{{\displaystyle |{\bf k}|^2 }}{{\displaystyle (k^3)^2}}
\left( x-\frac \xi 2\right)
M_N\left[ \left( x-\frac \xi 2\right) M_N-k^3\right]} \\
&+& \Biggl( \xi \to -\xi, \;\; \mbox{{\boldmath $\Delta$}}
\to -\mbox{{\boldmath $\Delta$}}
\Biggr)\, ,
\label{H-1-sym-res}
\ee
where the last integral over
transverse momenta is supposed to be regularized by the
Pauli--Villars method, {\em i.e.}, by subtracting the equivalent integral
with $M$ replaced by the regulator mass, $M_{PV}$; see \cite{DPPPW} for
details. [Here $\theta$ denotes the step function.]
Eq.~(\ref{H-1-sym-res}) has the remarkable property of being
discontinuous at $|x|=\xi/2$. Such behaviour
would lead immediately to a violation of the factorization theorems
for, say, deeply virtual Compton scattering (DVCS)
 processes\footnote{We are grateful to A.V.~Radyushkin for discussion
of this point}  \cite{Ji2},
since the expression for the DVCS amplitude contains a factor

\be
\mbox{Re}
\int_{-1}^1 dx \,  \left( \frac{1}{x-\xi/2+ i0}+\frac{1}{x+\xi/2+ i0}
\right) \, \xi \,  H(x,\xi,\Delta^2),
\label{dvcs}
\ee
which would be logarithmically divergent if $H(x,\xi,\Delta^2)$
were discontinuous at $|x|=\xi/2$. However, this conclusion is
premature: the discontinuities of eq.~(\ref{H-1-sym-res}) are
artifacts of neglecting the momentum dependence of the constituent
quark mass, as we shall now show.
\par
Let us analyze the original loop integral over $p$ in
eq.~(\ref{H-1-sym-res-mp}).
By inspection of the denominators it is easy to see that for $x$ close to
$\pm \xi/2$ the dominant contribution to the integral comes from
the region where the virtuality in one of the quark propagators
becomes large,
\be
(p-k)^2 &\sim& \frac{M_0(p_\perp^2+M_0^2)}{(|x|-\xi/2) M_N}\, .
 \label{pk}
\ee
However, in this region, due to the presence of the momentum--dependent
mass $M((p-k)^2)$ in the numerator, the whole expression drops to zero.
More precisely, one can see that for
\be
|x|-\xi/2 &\sim&  \frac{(M_0\rho)^2 M_0}{M_N}\, ,
\label{region}
\ee
where $\rho\sim (600\,{\rm MeV})^{-1}$ is the characteristic scale of
momentum dependence of the quark mass (the average instanton
size), the integral over $p_\perp$ is cut already at
$p_\perp^2\sim M_0^2$, whereas with a momentum--independent mass it
would always be cut at $p_\perp^2\sim 1/\rho^2$, see eq.~(\ref{H-1-sym-res}).
This analysis shows that the momentum dependence of the constituent
mass {\em can not} be neglected for $x$ in region eq.~(\ref{region}).
Consequently, keeping the momentum dependence of the constituent
quark mass, the discontinuities present in eq.~(\ref{H-1-sym-res}) are
smeared over an interval of order $(M_0 \rho)^2/N_c$, so that the
integral (\ref{dvcs}) is finite but it is still parametrically
(and numerically) enhanced by
a factor of $\log [(M_0\rho)^2 /N_c ]$; see the next section for
numerical estimates. For values of $x$ far from
$\pm \xi /2$ the momentum dependence of the mass can be
safely neglected and the simplified expression eq.~(\ref{H-1-sym-res})
gives reliable a approximation to eq.(\ref{H-1-sym-res-mp}).

Let us also note that the points $x=\pm \xi/2$ divide the interval
of the variable $x$ ($-1 \le x \le 1$) in three regions:
$x\le -\xi/2$, where the function  $H(x,\xi,\Delta^2)$ describes
the antiquark distribution; $x\ge \xi/2$, where it corresponds to
the quark distribution, and $-\xi\le x\le \xi/2$, where $H(x,\xi,\Delta^2)$
resembles a meson wave function. It is therefore natural that the
functions $H(x,\xi,\Delta^2)$ has crossovers at $|x|=\xi/2$. This
feature is remarkably reproduced in our model owing to the Dirac
continuum contribution. We note that in the forward limit,
$\Delta\to 0$, this crossover corresponds to the fact that {\it both}
the quark and antiquark distributions are {\it positive}; hence
the universal function $q(x)$ (which describes the quark distribution
at $x \ge 0$ and {\em minus} the antiquark distribution at $x\le0$)
evidently must exhibit a crossover at $x=0$.

\section{Numerical results and discussion}
\label{section_numeric}
We have calculated numerically the isosinglet distribution
$H(x,\xi,\Delta^2)$;
the analogous calculations for other OFQD's can be done along the same
lines. Our main purpose here is to discuss qualitative behaviour of the
OFQD's; to this end it is sufficient to consider $H(x,\xi,\Delta^2)$.
A comprehensive study of the other distributions will be given elsewhere.

For the numerical calculations we shall use the variational estimate of
the soliton profile, eq.(\eq{hedge}), of ref.\cite{DPP}
($M_0 = 350\, {\rm MeV}$),
\beq
P(r) \;\; = \;\; -2\;\arctan\left(\frac{r_0^2}{r^2}\right) ,
\hspace{1cm}
r_0 \;\; \approx \;\; 1.0/M_0 ,
\hspace{1cm} M_N \;\; \approx \;\; 1170\;{\rm MeV} ,
\la{varprof}
\eeq
which has been used in the calculation of usual parton distributions
in refs.\cite{DPPPW,DPPPW2}. Furthermore, we approximate the
momentum--dependent mass predicted by the instanton model of the
QCD vacuum \cite{DP} by the simple form
\be
M(-p^2) &=& \frac{M_0 \Lambda^{6}}{(\Lambda^2+p^2)^3},
\ee
where the parameter $\Lambda$ is related to the averaged instanton
size, $\rho$, by $\Lambda = 6^{1/3} \rho^{-1} $. This form reproduces
the asymptotic behaviour of $M(p^2)$ at large euclidean $p^2$
obtained in the instanton vacuum,
\be
\nonumber
M(-p^2) &\sim& \frac{36 M_0}{\rho^6 p^6} \hspace{2cm}
(p^2 \rightarrow \infty ).
\ee
We have explored also other forms of the momentum dependence of the mass
and found that numerically the results are very close to each other.

We estimate the Dirac continuum contribution to $H(x,\xi,\Delta^2)$
using the interpolation formula, eq.~(\ref{H-1-sym-res-mp}), which
gives a reliable approximation preserving all qualitative features
of the continuum contribution. The contribution of the discrete level
is calculated using eq.~(\ref{valence1}).

First we compute $H(x,\xi,\Delta^2)$ in the forward limit, $\Delta\to
0$, where it coincides with the usual quark and antiquark distributions.
The result is shown in Fig.~1, where we plot
separately the contributions
of the discrete level and that of Dirac continuum (computed from the
interpolation formula), as well as their sum.  It should be emphasized
that the distribution of {\em antiquarks} arising from the discrete
level (see \eq{valence1}) is definitely negative and sizeable.
\footnote{In the extreme case of a very strongly bound discrete level
when it approaches the lower continuum, this level would not produce
quarks at all -- only antiquarks, but with a negative sign!}.
Positivity of the parton distributions is restored only when one
includes the contribution of the Dirac continuum. The full
result for the function $H(x,0,0)$ (consisting of the level and continuum
contributions) exhibits strong crossover
at $x=0$, corresponding to the fact that {\em both} the quark and
antiquark distributions are positive. The crossover occurs in an
$x$-interval of order $|x| < \sim 0.05$; we shall see that this
remains so also for the off--forward case. To illustrate this we show
in Fig.~2 the distribution $H(x,\xi,\Delta^2)$ as a function of $x$,
for $\xi=0.3$ and $\Delta_T^2 \equiv -\Delta^2-\xi^2 M_N^2 = 0$.
Again we have plotted separately the contributions of the discrete level
and the Dirac continuum (according to the interpolation formula), as well
as their sum. We see that the discrete level contribution is a smooth
function which does not ``know'' about the points $x=\pm \xi/2$. This
kind of behaviour
has been assumed in all model approaches to OFQD's, {\it e.g.} in the bag
model calculations \cite{Ji3}. However the contribution of the Dirac
continuum changes drastically the picture: the function
$H(x,\xi,\Delta^2)$ shows now a fast crossover around the points
$x=\pm \xi/2$. The interval of $x$ over which this crossover occurs is
of the order of $\sim 0.05$, as in the forward case (see Fig.~1).
\par
The fast crossover of $H(x,\xi,\Delta^2)$ at $x=\pm \xi/2$.
may have physical implications. For example, it may lead to the considerable
enhancement of the DVCS amplitude, because the crossover occurs exactly at
the points where the integral eq.~(\ref{dvcs}) entering the DVCS
amplitude has singularities. Numerically, at $\xi = 0.3$ and
$\Delta_T^2 = 0$ the contribution of the Dirac continuum to the
integral (\ref{dvcs}) is 5.1 (almost the whole integral is collected
in small vicinity of point $x=\pm \xi/2$), while the corresponding
contribution of the discrete level is 3.2. We thus see that the
crossover in $H(x,\xi,\Delta)$ contributes more than 60\% to the integral
(\ref{dvcs}) and may thus considerably increase the
DVCS cross section.

In order to illustrate the dependence of $H(x,\xi,\Delta^2)$ on
$\xi$ and $\Delta^2$ we plot this function for a fixed momentum transfer
of $\Delta^2=-0.5$~GeV$^2$ for various values of $\xi$ (see Fig.~3),
and for fixed $\xi = 0.3$ and various values of momentum transfer
(see Fig.~4).

In the large $N_c$ limit the nucleon is heavy, so the OFPD's do not
automatically go to zero at $x = 1$; see the discussion in \cite{DPPPW}.
However, at $x\gg 1/N_c$ the distributions behave as
$\sim\exp(-\mbox{const}\cdot N_cx)$. Numerically, even for $N_c=3$ all
distributions computed are very small at $x\approx 1$.

\section{Conclusions}
\setcounter{equation}{0}
\label{section_summary}
In this paper we have investigated the off-forward quark distribution
(OFQD's) functions in the nucleon in the large--$N_c$ limit.
At large $N_c$ the nucleon can be viewed as a heavy
semiclassical body whose $N_c$ ``valence'' quarks are bound by a
self-consistent pion field. The energy of the pion field is given
by the effective chiral lagrangian and coincides with the aggregate
energy of the Dirac sea of quarks (the free continuum subtracted).
To compute the quark and antiquark distributions (forward and off--forward)
one must sum the contributions from the discrete level and from
the (distorted) negative-energy Dirac continuum.

We have found that the flavour singlet off--forward distribution
$H(x,\xi,\Delta^2)$ exhibits a qualitatively new behaviour
due to the contribution of the Dirac continuum: The function shows
fast crossovers around the points $x = \pm \xi/2$.
Our numerical estimates indicate that this crossover could lead to a
considerable enhancement of the deeply virtual Compton scattering
cross section.

In the crossover regions the behavior of the OFQD's is essentially
determined by the momentum dependence of the constituent quark mass
generated in the dynamical breaking of chiral symmetry. In particular,
the region in $x$ over which the crossover takes place is proportional
to $(M_0 \rho )^2/N_c$, where $\rho$ is the characteristic momentum scale
at which the dynamical quark mass drops to zero --- the average instanton
size. Thus, the enhancement of the deeply virtual Compton scattering
cross section is governed by the
small parameter intrinsic to the instanton model of the QCD vacuum,
the packing fraction of the instanton medium,
$(M_0 \rho )^2 \propto (\rho / R )^4$.

The OFQD calculated here in the chiral quark--soliton model
refers to a low normalization point of order the UV
cutoff intrinsic in this model, $\rho^{-1} \approx 600 \, {\rm MeV}$.
To obtain the OFQD's at higher normalization points one has to evolve
the ``primordial'' distributions using the evolution equations
derived in refs.\cite{Ji2,Rad2,evh,Rad3}. With regard to the
description of experiments it would be extremely interesting to
study to what extent the crossovers at $x = \pm \xi/2$ persist
when evolved to higher normalization points.
\\[1.2cm]
{\large\bf Acknowledgements}
\\[.3cm]
We are grateful to L.~Mankiewicz, M.~Strikman and
A.~Radyushkin for inspiring conversations. This work has been supported
in part by the RFBR (Moscow), Deutsche Forschungsgemeinschaft (Bonn)
and by COSY (J\"ulich). The Russian participants acknowledge the
hospitality of Bochum University.  M.V.P. is being supported by the
A.v.Humboldt Foundation.

\newpage
\appendix
\renewcommand{\theequation}{\Alph{section}.\arabic{equation}}
\section{Bound-state level contribution to $H(x,\xi,\Delta^2)$}
\setcounter{equation}{0}

We present here the contributions of the discrete bound-state
level to the singlet $H(x,\xi,\Delta^2)$.
The bound-state level occurs in the grand spin $K=0$ and parity
$\Pi=+$ sector of the Dirac hamiltonian (\ref{hU}). In that sector
the eigenvalue equation takes the form:

\be
\left(\begin{array}{cc}
 M \cos P(r) &
{\displaystyle -\frac{\partial}{\partial r}-\frac{2}{r} - M \sin P(r)}\\
{\displaystyle \frac{\partial}{\partial r} - M \sin P(r)} & - M \cos P(r)
\end{array}\right)
\left(\begin{array}{c}
h_0(r) \rule[-.5em]{0cm}{2em} \\ j_1(r) \rule[-.5em]{0cm}{2em}
\end{array}\right)
&=& E_{\rm lev}
\left(\begin{array}{c}
h_0(r) \rule[-.5em]{0cm}{2em} \\ j_1(r) \rule[-.5em]{0cm}{2em}
\end{array}\right).
\label{H-K-0:Pi-plus}
\ee
We assume that the radial wave functions are normalized by the
condition

\be
\int\limits_{0}^{\infty} dr \, r^2  \, [h_0^2(r) + j_1^2(r)] &=& 1.
\ee
We introduce the Fourier transforms of the radial wave functions,

\be
h(k) &=& \int\limits_{0}^{\infty} dr \, r^2\, h_0(r) R_{k0}(r),
\hspace{1.2cm}
j(k) \;\; = \;\; \int\limits_{0}^{\infty} dr \, r^2\, j_1(r) R_{k1}(r),
\ee
where
\be
R_{kl}(r) &=&  \sqrt{ \frac{k}{r}} J_{l+\frac12}(kr)
\;\; = \;\; (-1)^l\sqrt{\frac{2}{\pi}} \frac{r^l}{k^l}
\left( \frac{1 }{r } \frac{d}{dr } \right)^l \frac{\sin kr }{r}.
\ee

The bound-state level contribution to the singlet $H(x,\xi,\Delta^2)$
distribution function can be simply obtained from the general
eq.~(\ref{H-singlet-general}). We get:

\be
\nonumber
&&\sum\limits_fH_f(x,\xi ,\Delta ^2)_{\rm lev}=
2\pi N_c M_N \int \frac{d^2{\bf k_\perp}}{(2\pi)^2}
\frac{1}{k k'}\Biggl\{
h(k)h(k') \\
\label{valence1}
&+&j(k)j(k')
\frac{{\bf k_\perp}\cdot({\bf k_\perp}+
\mbox{{\boldmath $\Delta_\perp$}})+
(xM_N-E_{\rm lev})^2-\frac 14 \xi^2 M_N^2}{k k'}\\
\nonumber
&-&h(k)j(k') \frac{x M_N-E_{\rm lev}+\frac 12 \xi M_N}{k}
-h(k')j(k) \frac{x M_N-E_{\rm lev}-\frac 12 \xi M_N}{k'}
\Biggr\} \, ,
\ee
where
\be
k&=&\sqrt{{\bf k_\perp}^2+ ((x+\frac 12 \xi)M_N-E_{\rm lev})^2}\\
k'&=&\sqrt{({\bf k_\perp+
\mbox{{\boldmath $\Delta_T$}}})^2+
\bigl[(x-\frac 12 \xi)M_N-E_{\rm
lev}\bigr]^2}\, .
\ee
Note that the r.h.s. of eq.~(\ref{valence1})
is positive in the forward limit ($\Delta \to 0$)
for all values of $x$, in particular
at $x<0$ where eq.~(\ref{valence1}) determines in fact the
antiquark distribution.  Since $\bar q(x)=-H(-x,0,0)$, it means that
eq.~(\ref{valence1}) gives a {\em negative} distribution of antiquarks
at $x>0$. At the same time it is easy to check by integrating over $x$
eq.~(\ref{valence1}) that the baryon number sum rule is fully saturated
by the discrete-level contribution only.

\newpage
\begin{figure}
 \vspace{-1cm}
\epsfxsize=16cm
\epsfysize=15cm
\centerline{\epsffile{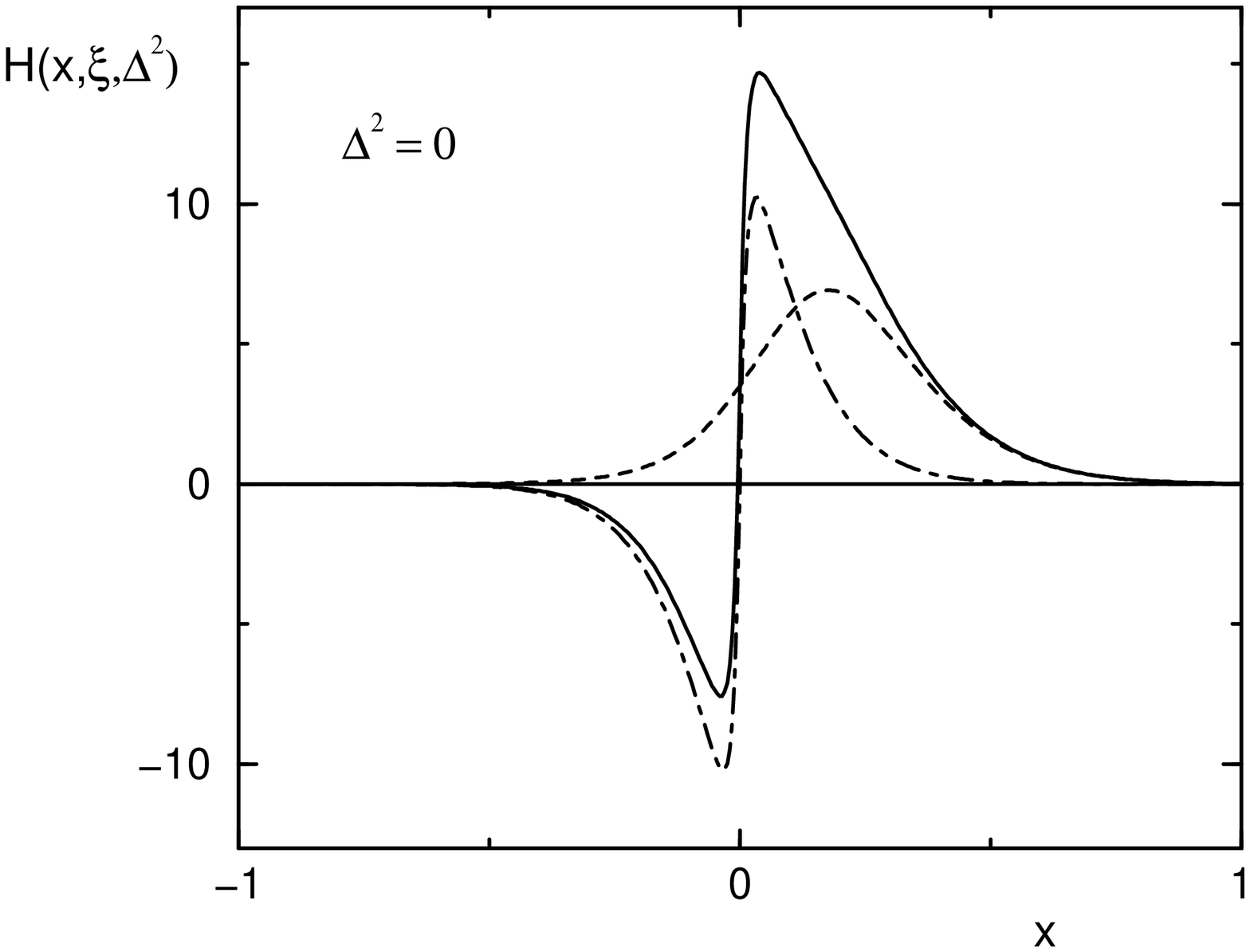}}
\caption[]{ The isosinglet distribution $H(x,\xi,\Delta^2)$ in the forward
limit, $\Delta=0$.
{\em Dashed line}:
contribution from the discrete level. {\em Dashed-dotted line}:
contribution from the Dirac continuum according to the interpolation
formula, eq.~(\protect\ref{H-1-sym-res-mp}).
{\em Solid line}: total distribution (sum of the dashed
and dashed-dotted curves).} \end{figure}

\newpage
\begin{figure}
 \vspace{-1cm}
\epsfxsize=16cm
\epsfysize=15cm
\centerline{\epsffile{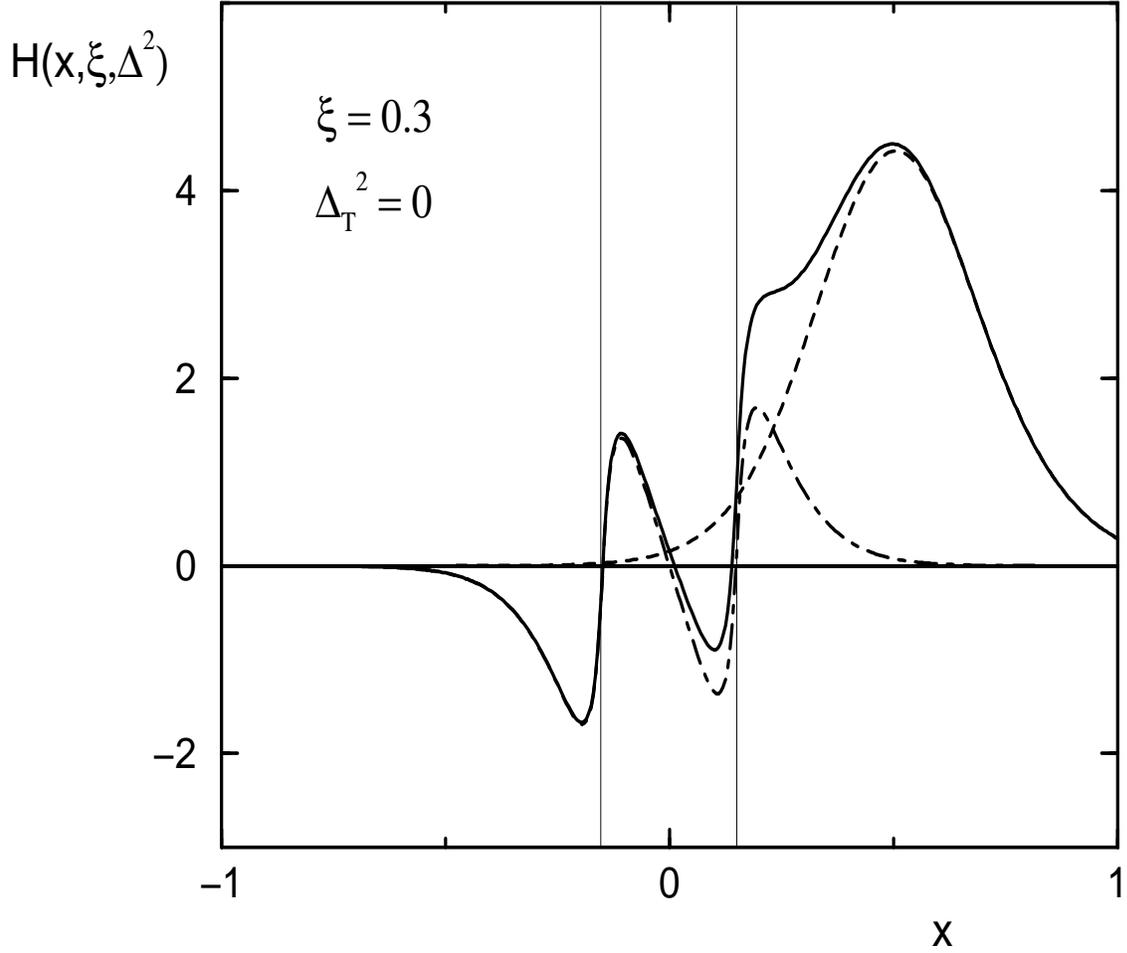}}
\caption[]{The isosinglet distribution $H(x,\xi,\Delta^2)$ for
$\Delta_T^2=0$ ($\Delta_T^2 \equiv -\Delta^2-\xi^2 M_N^2$) and $\xi=0.3$.
{\em Dashed line}:
contribution from the discrete level. {\em Dashed-dotted line}:
contribution from the Dirac continuum according to the interpolation
formula, eq.~(\protect\ref{H-1-sym-res-mp}).
{\em Solid line}: the total distribution (sum of the dashed
and dashed-dotted curves). The vertical lines mark the crossover points
$x=\pm \xi/2$.
} \end{figure}

\newpage
\begin{figure}
 \vspace{-1cm}
\epsfxsize=16cm
\epsfysize=15cm
\centerline{\epsffile{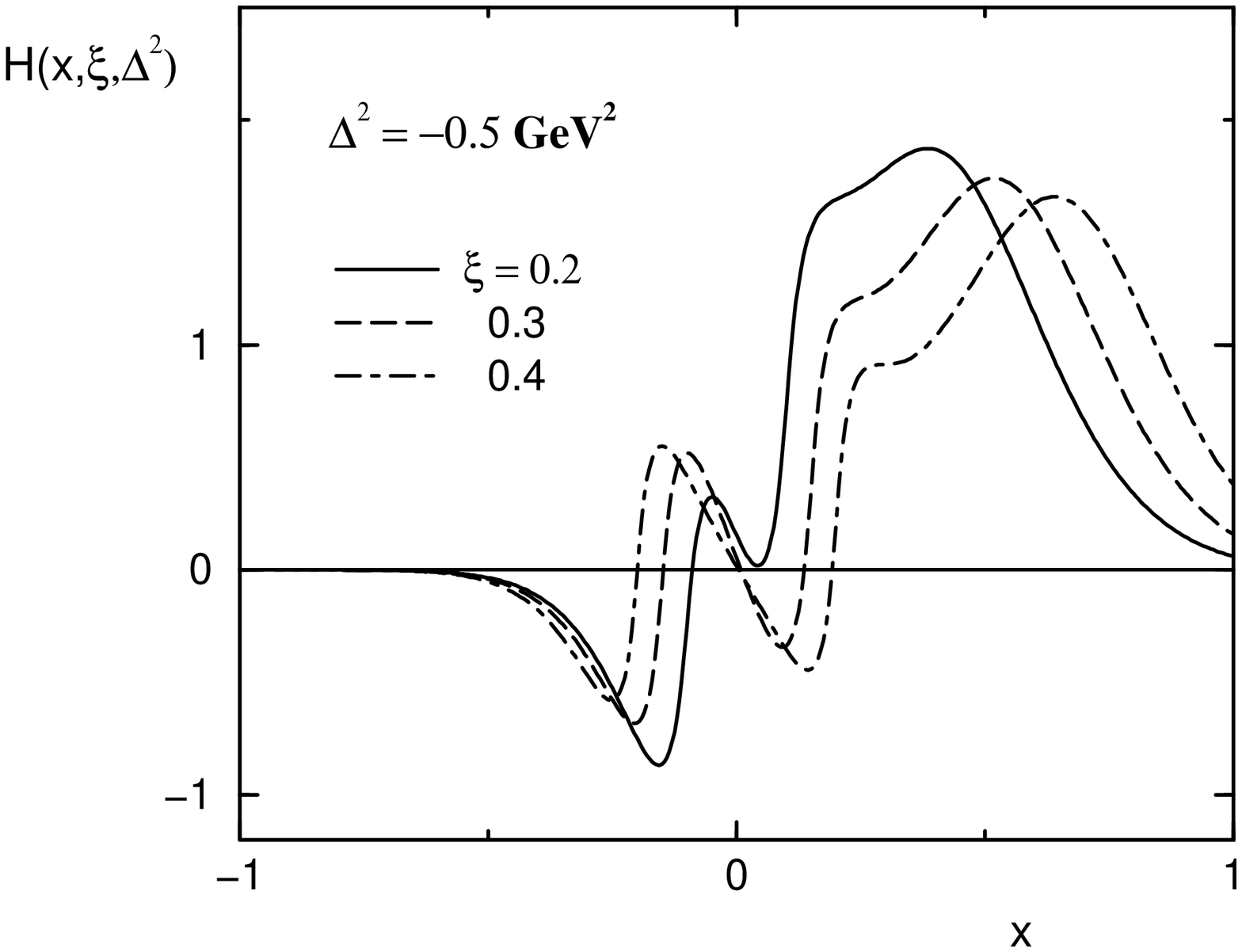}}
\caption[]{The isosinglet distribution $H(x,\xi,\Delta^2)$ (total result)
for fixed $\Delta^2=-0.5$~GeV$^2$ and various values of $\xi$.
} \end{figure}

\newpage
\begin{figure}
 \vspace{-1cm}
\epsfxsize=16cm
\epsfysize=15cm
\centerline{\epsffile{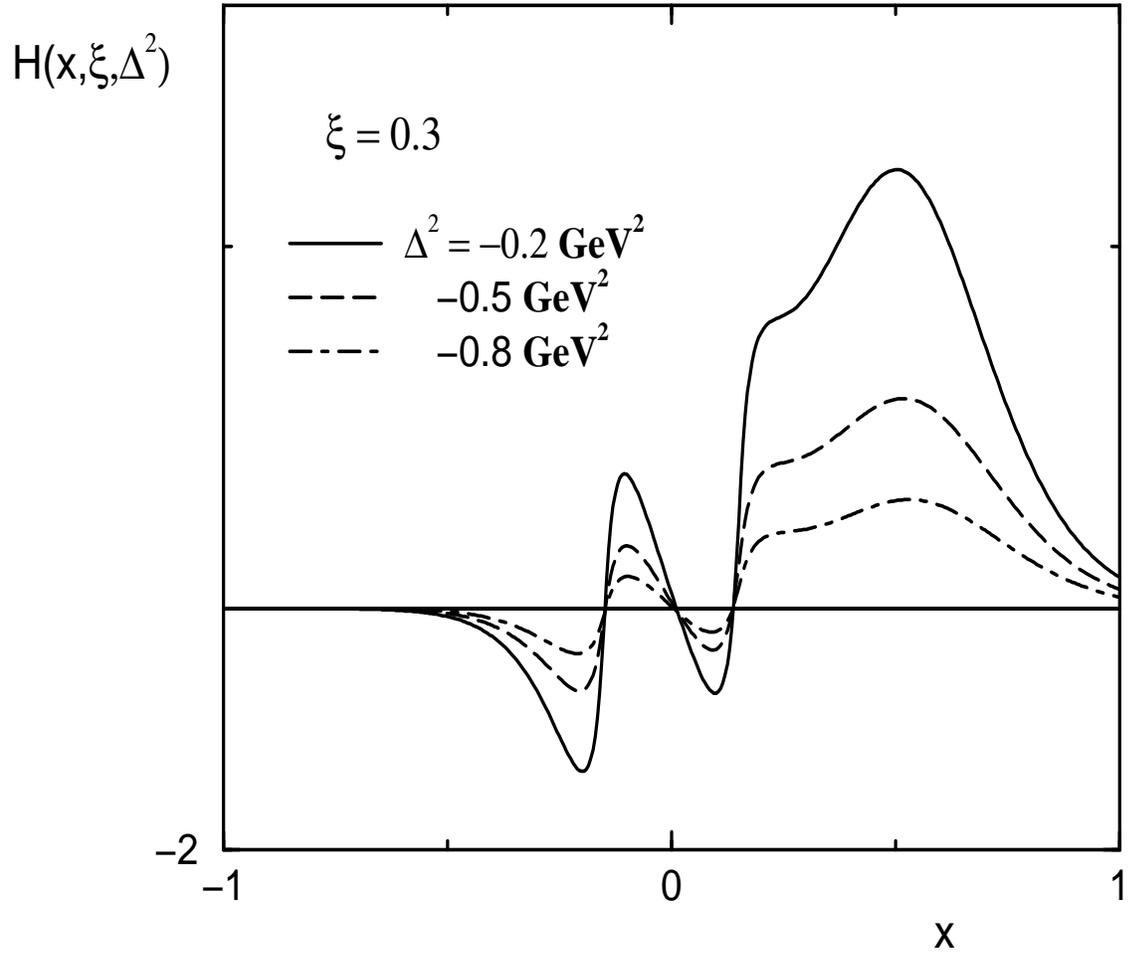}}
\caption[]{The isosinglet distribution $H(x,\xi,\Delta^2)$ (total result)
for fixed $\xi=0.3$ and various values of $\Delta^2$.
} \end{figure}

\end{document}